\documentclass[sigconf,screen]{acmart}

\AtBeginDocument{%
  }

\copyrightyear{2025}
\acmYear{2025}
\setcopyright{cc}
\setcctype{by}
\acmConference[BDCAT '25]{IEEE/ACM 12th International Conference on Big Data Computing, Applications and Technologies}{December 1--4, 2025}{Nantes, France}
\acmBooktitle{IEEE/ACM 12th International Conference on Big Data Computing, Applications and Technologies (BDCAT '25), December 1--4, 2025, Nantes, France}\acmDOI{10.1145/3773276.3774299}
\acmISBN{979-8-4007-2286-8/2025/12}

\usepackage{algpseudocode}
\usepackage{algorithm}
\usepackage[inline,shortlabels]{enumitem}
\usepackage[capitalise]{cleveref}
\usepackage{pgfplots}
\usepackage{pgfplotstable}
\usepackage{booktabs}
\usepackage[table]{xcolor} %
\usepackage{tikz}
\usepackage{subcaption}
\usepackage{forest}
\usepackage{balance}
\usepgfplotslibrary{dateplot,statistics,groupplots}
\pgfplotsset{compat=1.18}

\pgfplotsset{
    box plot/.style={
        /pgfplots/.cd,
        black,
        only marks,
        mark=-,
        mark size=\pgfkeysvalueof{/pgfplots/box plot width},
        /pgfplots/error bars/y dir=plus,
        /pgfplots/error bars/y explicit,
        /pgfplots/table/x index=\pgfkeysvalueof{/pgfplots/box plot x index},
    },
    box plot box/.style={
        /pgfplots/error bars/draw error bar/.code 2 args={%
            \draw  ##1 -- ++(\pgfkeysvalueof{/pgfplots/box plot width},0pt) |- ##2 -- ++(-\pgfkeysvalueof{/pgfplots/box plot width},0pt) |- ##1 -- cycle;
        },
        /pgfplots/table/.cd,
        y index=\pgfkeysvalueof{/pgfplots/box plot box top index},
        y error expr={
            \thisrowno{\pgfkeysvalueof{/pgfplots/box plot box bottom index}}
            - \thisrowno{\pgfkeysvalueof{/pgfplots/box plot box top index}}
        },
        /pgfplots/box plot
    },
    box plot top whisker/.style={
        /pgfplots/error bars/draw error bar/.code 2 args={%
            \pgfkeysgetvalue{/pgfplots/error bars/error mark}%
            {\pgfplotserrorbarsmark}%
            \pgfkeysgetvalue{/pgfplots/error bars/error mark options}%
            {\pgfplotserrorbarsmarkopts}%
            \path ##1 -- ##2;
        },
        /pgfplots/table/.cd,
        y index=\pgfkeysvalueof{/pgfplots/box plot whisker top index},
        y error expr={
            \thisrowno{\pgfkeysvalueof{/pgfplots/box plot box top index}}
            - \thisrowno{\pgfkeysvalueof{/pgfplots/box plot whisker top index}}
        },
        /pgfplots/box plot
    },
    box plot bottom whisker/.style={
        /pgfplots/error bars/draw error bar/.code 2 args={%
            \pgfkeysgetvalue{/pgfplots/error bars/error mark}%
            {\pgfplotserrorbarsmark}%
            \pgfkeysgetvalue{/pgfplots/error bars/error mark options}%
            {\pgfplotserrorbarsmarkopts}%
            \path ##1 -- ##2;
        },
        /pgfplots/table/.cd,
        y index=\pgfkeysvalueof{/pgfplots/box plot whisker bottom index},
        y error expr={
            \thisrowno{\pgfkeysvalueof{/pgfplots/box plot box bottom index}}
            - \thisrowno{\pgfkeysvalueof{/pgfplots/box plot whisker bottom index}}
        },
        /pgfplots/box plot
    },
    box plot median/.style={
        /pgfplots/box plot,
        /pgfplots/table/y index=\pgfkeysvalueof{/pgfplots/box plot median index}
    },
    box plot width/.initial=1em,
    box plot x index/.initial=0,
    box plot median index/.initial=1,
    box plot box top index/.initial=2,
    box plot box bottom index/.initial=3,
    box plot whisker top index/.initial=4,
    box plot whisker bottom index/.initial=5,
}

\begin{document}

\title{Gaia: Hybrid Hardware Acceleration for Serverless AI in the 3D Compute Continuum}


\author{Maximilian Reisecker}
\orcid{0009-0001-3354-0981}
\affiliation{%
  \institution{Distributed Systems Group}
  \city{TU Wien, Vienna}
  \country{Austria}}
\email{e11908094@student.tuwien.ac.at}

\author{Cynthia Marcelino}
\orcid{0000-0003-1707-3014}
\affiliation{%
  \institution{Distributed Systems Group}
  \city{TU Wien, Vienna}
  \country{Austria}}
\email{c.marcelino@dsg.tuwien.ac.at}

\author{Thomas Pusztai}
\orcid{0000-0001-9765-6310}
\affiliation{%
  \institution{Distributed Systems Group}
  \city{TU Wien, Vienna}
  \country{Austria}}
\email{t.pusztai@dsg.tuwien.ac.at}

\author{Stefan Nastic}
\orcid{0000-0003-0410-6315}
\affiliation{%
  \institution{Distributed Systems Group}
  \city{TU Wien, Vienna}
  \country{Austria}}
\email{snastic@dsg.tuwien.ac.at}


\begin{abstract}
  Serverless computing offers elastic scaling and pay-per-use execution, making it well-suited for AI workloads. As these workloads run in heterogeneous environments such as the Edge-Cloud-Space 3D Continuum, they often require intensive parallel computation, which GPUs can perform far more efficiently than CPUs. 
  However, current platforms struggle to manage hardware acceleration effectively, as static user-device assignments fail to ensure SLO compliance under varying loads or placements, and one-time dynamic selections often lead to suboptimal or cost-inefficient configurations.
  
  To address these issues, we present Gaia, a GPU-as-a-service model and architecture that makes hardware acceleration a platform concern. Gaia combines (i) a lightweight Execution Mode Identifier that inspects function code at deploy time to emit one of four execution modes, and a Dynamic Function Runtime that continuously re-evaluates user-defined SLOs to promote or demote between CPU- and GPU backends.
  Our evaluation shows that it seamlessly selects the best hardware acceleration for the workload, reducing end-to-end latency by up to 95\%. These results indicate that Gaia enables SLO-aware, cost-efficient acceleration for serverless AI across heterogeneous environments.
  
\end{abstract}



\maketitle

\section{Introduction}

Serverless computing offers elastic scaling, and a pay-per-use model, making it well-suited for a wide range of applications such  as real-time data processing, IoT, and AI
inference~\cite{AdvancingServerlessAI2024,optimus2024,ServerlessLLM}. 
However, hardware acceleration in serverless platforms remains challenging. Functions are typically deployed without knowledge of the underlying hardware, while developers must manually decide whether a function should run on a CPU or GPU. 
The static assignment of functions to hardware resources increases developer complexity and often results in inefficient utilization of heterogeneous environments, where resources may be unavailable for function requirements. 

Recently, the rapid growth of Low Earth Orbit (LEO) satellite megaconstellations equipped with onboard computing resources and inter-satellite links (ISLs) has enabled novel paradigms in space, such as serverless computing~\cite{Krios2024,MillionSats} in the 3D Compute Continuum~\cite{HyperDrive2024,cosmos,gravara2025novel}. The 3D Compute Continuum unifies the Edge, Cloud, and LEO satellites into one system, creating a seamless application flow. 
In such environments, efficient hardware acceleration becomes critical. 
In particular, LEO satellites introduce challenges that amplify the problem, such as resource availability as satellites move in and out of connectivity windows, constrained hardware capacity, and network conditions that fluctuate with orbital dynamics~\cite{Network27k,EDRS_Overview,DelayNoOption}. 
Therefore, a function assigned to a GPU during deployment time may not find GPU capacity on a newly scheduled satellite node, causing scheduling delays or failures. Moreover, one-off dynamic decisions may allocate GPUs unnecessarily during temporary workload peaks, leading to severe cost inefficiencies in an environment where resources are scarce and expensive.  
As a result, functions may suffer from performance degradation under workload drift or incur increased costs due to overprovisioned GPU usage. 

%Moreover, existing dynamic solutions typically make one-off device choices at deployment or invocation time, but fail to revisit them as workloads and system conditions change. 

Common approaches that enable serverless hardware acceleration include:
\begin{enumerate*} [label=(\alph*)]
\item \emph{Static} state-of-the-art \cite{kim_gpu_2018,pemberton_kernel-as--service_2022, naranjo_accelerated_2020} and commercial \cite{runpod,cerebrium,beam-cloud} approaches require the developer to decide whether to run on a CPU or GPU. While this provides fine-grained control, it increases developer complexity and inefficient resource allocation, as a platform in a heterogeneous environment might not have the capacity to execute the required hardware specifications. This leads to delays and potential errors.
\item \emph{Dynamic} approaches migrate to GPUs based on the current function performance \cite{awsbatch, bhasi_towards_2024} or determine it before execution, based on resource availability and user-defined SLOs such as performance, and cost \cite{romero_llama_2021,fingler_dgsf_2022}. However, most dynamic approaches make a one-off device choice and do not revisit it as load changes, which can lead to performance degradation under workload drift or increased costs.
\end{enumerate*}

Existing state-of-the-art approaches that enable hardware acceleration, i.e., GPU serverless function execution, either rely on static developer choices or one-off dynamic device selection~\cite{kim_gpu_2018,RAUSCH2021259}. Therefore, they fail to adapt to heterogeneous and dynamic environments such as the 3D Compute Continuum. As satellites move in and out of orbit and the connectivity windows shift, the platform must continuously reevaluate node capacity and available hardware resources. To address this gap, we present Gaia, an abstraction for seamless hardware acceleration in serverless platforms. Gaia enables the platform to automatically identify whether a function requires GPU or CPU execution and to adjust execution dynamically at runtime, while ensuring SLO compliance.
Our contributions include:

\begin{itemize}
    \item \emph{Gaia} a hybrid acceleration model and architecture that abstracts function requirements and enables seamless runtime adjustment between CPU and GPU, while complying with SLOs.
    \item \emph{Execution Mode Identifier} that identifies functions with GPU-affine operations at deployment time, providing hints to the serverless platform without requiring developer intervention, such as configuration or code changes.
    \item \emph{Dynamic Function Runtime Reconfiguration} that dynamically adjusts hardware allocation during execution as workloads or system conditions change, moving task execution to another node if necessary, thus preventing performance degradation and cost inefficiencies inherent to one-off decisions.
\end{itemize}

This paper has eight sections. 
\cref{motivation} presents the illustrative scenario and research questions.
\cref{architecture} describes the Gaia model as well as the architecture overview. 
\cref{runtime} describes the static hardware identifier and the dynamic runtime adaptation introduced by Gaia and their usage. 
\cref{impl} shows the prototype implementation details.
\cref{evaluation} discusses the experiments and evaluation, 
\cref{related_work} presents related work. 
\cref{conclusion} concludes with a final discussion and future work.

\section{Illustrative Scenario and Research Challenges}\label{motivation}
To better understand the challenges associated with hardware acceleration execution, in this section, we present our motivating illustrative scenario and our research challenges.

\subsection{Illustrative Scenario}\label{subsec:scenario}

\begin{figure}
    \centering
    \includegraphics[width=\linewidth]{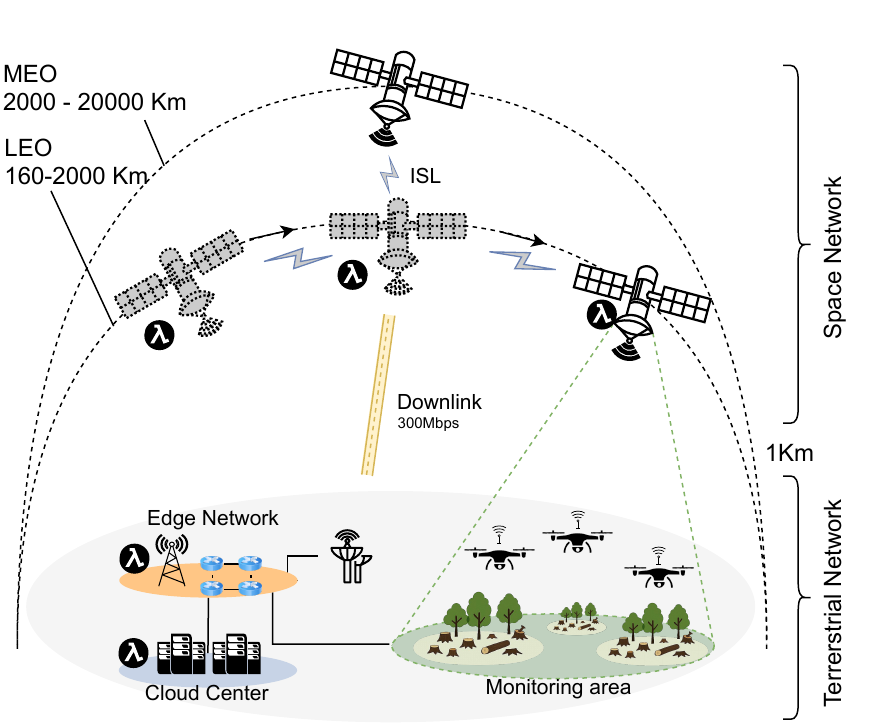}
    \caption{Deforestation Illustrative Scenario with Serverless Functions Execution in the 3D Compute Continuum}
    \label{fig:scenario}
\end{figure}

Early detection of deforestation in remote regions requires timely data processing and efficient use of heterogeneous resources across the 3D Continuum. In our deforestation use case (\cref{fig:scenario}), ground-based sensors and drones collect environmental indicators such as high-resolution imagery, temperature, and CO\textsubscript{2} levels. When in range, drones offload data to nearby edge nodes; otherwise, they transmit data directly to LEO satellites, ensuring high availability even when the edge network is unavailable. Moreover, LEO satellites can also fuse ground data from drones with imagery from Earth Observation (EO) satellites, including optical and infrared feeds. The combined streams are processed through a serverless workflow consisting of ingestion, image segmentation, and pattern recognition functions to detect deforestation hotspots. By distributing these tasks across edge, cloud, and space, the 3D Continuum reduces data transfer overheads, lowers detection latency, and supports scalable analysis of daily large EO datasets up to 1,5TB that would otherwise be costly to transmit to ground-based cloud services~\cite{ESA_Sentinel2Ops, Sentinel2CLaunched2024}.
Gaia enhances this workflow by abstracting CPU/GPU hardware selection, ensuring that resource-intensive functions, such as Object Detection or Prepare Dataset, seamlessly exploit available GPUs at Edge nodes or onboard LEO satellites, without requiring developers to manually specify execution devices. 
The Gaia static code analyzer identifies GPU-affine functions at deployment time, providing execution hints that prevent unnecessary GPU allocation for lightweight functions such as Ingest. Furthermore, Gaia’s adaptive runtime reconfiguration allows the platform to dynamically reassign functions when satellites move in and out of orbit or when workload intensity changes, maintaining SLO compliance despite fluctuating resource availability. Thus, Gaia enables the deforestation detection workflow to leverage onboard GPUs when ground links are congested, or to fall back to CPUs when GPU capacity is limited or unavailable, optimizing performance and cost while ensuring uninterrupted serverless execution.

\subsection{Research Challenges}

\emph{RC-1: How can serverless AI functions maintain optimal execution performance under resource constraints and dynamic conditions of the 3D Compute Continuum?}

The 3D Compute Continuum paradigm unifies Edge-Cloud and LEO satellites by leveraging the unique strengths of each layer.
Onboard processing is particularly beneficial for downstream applications such as Earth Observation analytics, as it avoids costly and delay-prone downlinking that depends on orbital passes, typically incurring multiple hours~\cite{L2D2_2021, SentinelTimeliness2025}.
Downstream applications often involve AI workloads that require hardware acceleration, such as GPUs. As satellites move in and out of range, certain functions must remain Earth-anchored to geo-specific monitoring regions, requiring migration to different satellites and adaptation of execution to maintain SLO compliance.
While GPUs are becoming more common on satellites~\cite{GpuAtSat2023, SatelliteCompCOTS2024}, their capacities are still limited by power constraints~\cite{EdgeComputingVisions6G2024, SmallSatellitesSurvey2018}, thermal constraints, and radiation hardening requirements~\cite{SpaceEnvEffectsNASA2020, SatelliteComputingCloudNative2023, SatelliteComputingVision2023}.
Thus, it is imperative to use the available capacities wisely.

\emph{RC-2: How to seamlessly identify functions' hardware requirements, such as GPU?}

Identifying the hardware requirements of a serverless function is an important, yet challenging, task to ensure that its end-to-end runtime SLO is met and that costs are limited~\cite{ServerlessComputingArchitectureSurvey2022, ServerlessSurveyIEEE_TranServComp2023}.
For CPU and memory requirements, there are various approaches to automate this through offline profiling~\cite{MAFF2022, CPU_TAMS2022, SLAM2022, StepConfJournal2024, ChunkFunc2025} or through online optimization that adjusts function resources based on previous results~\cite{Sizeless2021, Astra2021, Orion2022, FaaSDeliver2023}.
For GPUs, serverless functions commonly use either a granular sharing approach, where an entire GPU is allocated to a function~\cite{Hydrozoa2022, Elasticflow2023, ServerlessLLM}, or spatial partitioning~\cite{GSlice2020, INFless2022, FaST_GShare2023}. However, running kernels cannot be interrupted; hence, allocations persist for their execution duration.
To avoid having too few resources, many systems tend to overprovision GPU resources, thus leading to wasted capacities~\cite{Dilu2025}, something that must be avoided on satellites or Edge devices.

\emph{RC-3: How to dynamically change function runtime between CPU and GPU to balance performance and cost SLOs?}

Typically, serverless functions rely on the runtime that bridges the communication between the hardware and the function~\cite{PerformanceIsolation2025,faasm}.
Nevertheless, developers typically have to statically select whether the function should use a GPU-enabled runtime (which has a longer cold start delay) or a CPU-only runtime~\cite{kim_gpu_2018, naranjo_accelerated_2020,  pemberton_kernel-as--service_2022}.
This may result in both overprovisioning of GPU resources or limiting functions to the CPU, even though they could benefit from a GPU.
A simple approach to make this decision dynamic is to run a function on the CPU first and move it to the GPU if it violates the end-to-end SLO~\cite{GPU_ServerlessMultimedia2021}.
A more complex approach to dynamicity is to predict a function's execution time and to dynamically allocate it to the CPU or GPU depending on the end-to-end SLO and cost constraints~\cite{romero_llama_2021}.
However, estimating the execution time of a function is not trivial, and running this decision logic for every invocation may add considerable overhead.

\section{Gaia Design Principles and Architecture}\label{architecture}

In this section, we outline Gaia’s design principles and describe the architecture that implements them.

\subsection{Gaia Design Principles}

Gaia is designed to make hardware acceleration a platform concern while keeping functions portable and reducing developer effort. Therefore, we follow the principles that guided our architecture and are reflected in the mechanisms described in \cref{runtime,impl}.

\paragraph{Hardware-agnostic} Functions are authored without binding to hardware acceleration devices; execution mode is inferred at deploy time and adjusted at runtime.

\paragraph{Zero Developer Friction} No code changes are required; developers can opt into \texttt{auto (Gaia)|cpu|gpu} or rely on the execution mode identifier to transparently select the best hardware acceleration.

\paragraph{Intelligent Start \& Dynamic Switching.} Upon deployment of a function, Gaia performs a static analysis to determine the best-suited initial configuration. This decision is periodically re-evaluated at runtime, causing a dynamic switch to a different execution mode if necessary.

\paragraph{Observability by Design.} Each runtime decision is backed by telemetry, which monitors SLO requirements such as latency and throughput, and persists rationale (last-mode, measured latencies).

% \paragraph{Cost optimization.} While SLOs have priority, Gaia will attempt to minimize the costs if an SLO allows it.

\subsection{Gaia Architecture}

Gaia leverages static code features to identify hardware execution mode and monitoring to evaluate dynamic runtime changes for serverless functions across heterogeneous hardware acceleration backends with minimal developer burden. Functions are deployed with an initial execution mode inferred offline by the static code analyzer; at runtime, Gaia continuously evaluates SLO metrics to promote or demote the function’s backend, aligning performance and cost with SLOs across the 3D Compute Continuum.

\begin{figure}[t]
    \centering
    \includegraphics[width=\linewidth]{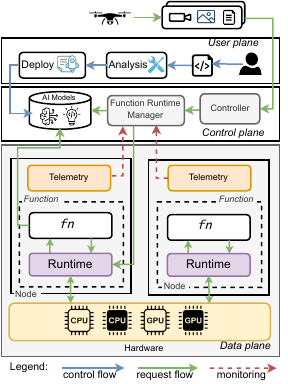}
    \caption{Gaia Architecture}
    \label{fig:architecture}
\end{figure}

The Gaia architecture, shown in \cref{fig:architecture}, is organized into three planes: user, control, and data.
In the user plane, the Static Code Analyzer classifies functions and annotates the manifest with execution hints.
In the control plane, a coordinator performs dynamic runtime management, closing the loop with telemetry to enact safe, seamless mode switches.
In the data plane, the function code and runtime execute, interfacing with the underlying hosting stack to utilize proper hardware acceleration, such as CPU or GPU backends, as required.

\subsubsection{Gaia Components}
Gaia is composed of six main components: Code Analyzer, Build and Deploy, Controller, Function Runtime Manager, Function Runtime, and Telemetry.

\paragraph{Code Analyzer} The Code Analyzer inspects serverless functions before deployment time. It parses the function code into an abstract syntax tree (AST) to identify GPU-related imports, explicit device calls, and tensor operations. Based on these heuristics, it assigns an execution mode (\texttt{cpu}, \texttt{cpu\_preferred}, \texttt{gpu\_preferred}, \texttt{gpu}) and annotates the function manifest accordingly. This ensures that the serverless platform schedules functions on suitable nodes without requiring changes to user code.  

\paragraph{Build and Deploy} Gaia enables three different deployment modes, allowing developers to explicitly specify whether a function should run on CPU, GPU, or use Gaia’s adaptive mode. When the adaptive mode is selected, the Code Analyzer is invoked during build and deployment, and its results are embedded in function deployment for the serverless platform.  

\paragraph{Controller} It manages the lifecycle of functions, and how functions are exposed, scaled, and updated. It manages incoming requests and distributes them across available instances, adjusts the number of active instances based on demand, and orchestrates updates so that newer versions of a function can replace older ones without interrupting ongoing requests. It is also responsible for scheduling functions on nodes with hardware acceleration specification received from the Function Runtime Manager.

\paragraph{Function Runtime Manager} It watches deployed functions and applies function runtime mode switches seamlessly. It interacts with the Controller bz giving hardware acceleration hints to dynamically adjust functions when performance and cost SLOs are exceeded. Dynamic shifting between CPU and GPU ensures efficient use of heterogeneous hardware while minimizing disruption to ongoing function executions.  

\paragraph{Function Runtime} Gaia supports multiple execution backends via container shims. The default container shim executes functions on CPU, while the NVIDIA container runtime shim enables GPU acceleration. Switching execution mode is achieved by redeploying the function with the appropriate shim, allowing Gaia to leverage GPU resources without modifying user workloads.

\paragraph{Telemetry} It component collects runtime metrics such as request latency, throughput, and resource utilization. These statistics are stored and continuously monitored by the Function Runtime Manager. Telemetry provides the feedback loop that allows Gaia to make informed adaptation decisions, ensuring that execution mode changes are based on actual performance data.

\section{Gaia Mechanisms}\label{runtime}

In this section, we present Gaia mechanisms for adapting serverless functions across CPU and GPU: a static Execution Mode Identifier that infers an initial mode from source code, and a Dynamic Function Runtime that monitors latency and request rate to promote or demote to minimize latency and cost and meet SLOs.

\subsection{Execution Mode Identifier}
The Execution Mode Identifier provides the first decision step in adapting a function’s runtime. Its purpose is to analyze the static code of a serverless function before deployment and assign it to a suitable execution mode. By examining imports, explicit hardware calls, and computational patterns, the component predicts whether the function is best executed on CPU, requires GPU acceleration, or can operate on either with a preference. 
This static classification reduces unnecessary runtime overhead, ensures clear cases (e.g., explicit GPU usage) are handled deterministically, and leaves more ambiguous cases to be fine-tuned later by the dynamic runtime adaptation mechanism.  

The Execution Mode Identifier analyzes the source code of a serverless function to assign one of four execution modes: 
\texttt{cpu}, \texttt{gpu}, \texttt{cpu\_preferred}, and \texttt{gpu\_preferred}. 

\begin{algorithm}[t]
\caption{Execution Mode Identifier}
\label{alg:static-simplified}
\begin{algorithmic}[1]

\Require Function $f$
\Ensure $m \in \{\texttt{CPU}, \texttt{CPU\_PREF}, \texttt{GPU\_PREF}, \texttt{GPU}\}$, reason $r$

\State Parse $f$ into AST \label{ln:Alg1_1}
\State Initialize flags: \texttt{dl\_import} $\gets$ \textbf{false}, \texttt{gpu\_explicit} $\gets$ \textbf{false}, 
       \texttt{big\_ops} $\gets$ \textbf{false}, \texttt{small\_ops} $\gets$ \textbf{false}  \label{ln:Alg1_2}

\ForAll{AST nodes} \label{ln:Alg1_3}
  \If{import of \texttt{torch} or \texttt{tensorflow}} 
    \State \texttt{dl\_import} $\gets$ \textbf{true} \label{ln:Alg1_4}
  \ElsIf{GPU call like \texttt{to("cuda")}, \texttt{.cuda()}, \texttt{torch.device("cuda")} 
         \textbf{and not} \texttt{cuda.is\_available()}}
    \State \texttt{gpu\_explicit} $\gets$ \textbf{true} \label{ln:Alg1_5}
  \ElsIf{tensor operation}
    \State Estimate tensor size; set \texttt{big\_ops} or \texttt{small\_ops}  \label{ln:Alg1_6}
  \EndIf
\EndFor

\If{\texttt{gpu\_explicit}}  \label{ln:Alg1_7}
  \State $m \gets \texttt{GPU}$, $r \gets$ ``explicit GPU usage''
\ElsIf{\texttt{dl\_import} \textbf{and} \texttt{big\_ops}}
  \State $m \gets \texttt{GPU\_PREF}$, $r \gets$ ``large tensor ops''
\ElsIf{\texttt{dl\_import} \textbf{and} \texttt{small\_ops} \textbf{and not} \texttt{big\_ops}}
  \State $m \gets \texttt{CPU\_PREF}$, $r \gets$ ``small tensor ops''
\ElsIf{\texttt{dl\_import}}
  \State $m \gets \texttt{CPU\_PREF}$, $r \gets$ ``imports only''
\Else
  \State $m \gets \texttt{CPU}$, $r \gets$ ``no GPU-related activity''
\EndIf

\State \Return $(m, r)$  \label{ln:Alg1_8}

\end{algorithmic}
\end{algorithm}

The function code is first parsed into an AST (line \ref{ln:Alg1_1}), and flags are initialized to record relevant patterns (line~\ref{ln:Alg1_2}). 
The AST is then traversed node by node (line~\ref{ln:Alg1_3}). Imports of deep learning frameworks such as PyTorch or TensorFlow set the \texttt{dl\_import} flag (line~\ref{ln:Alg1_4}). 
Explicit GPU calls such as \texttt{torch.device("cuda")} or \texttt{.to("cuda")} set the \texttt{gpu\_explicit} flag (line~ \ref{ln:Alg1_5}). 
Tensor operations are analyzed to estimate workload size; depending on the result, either \texttt{big\_ops} or \texttt{small\_ops} is marked (line~ \ref{ln:Alg1_6}). 

At decision time (starting from line~\label{ln:Alg1_7}), the algorithm follows a hierarchical process. 
Explicit GPU calls enforce the \texttt{gpu} mode. 
Large tensor operations, combined with deep learning imports, lead to \texttt{gpu\_preferred}. 
Small tensor operations or imports without heavy usage result in \texttt{cpu\_preferred}. 
When no GPU-related activity is detected, the execution mode defaults to \texttt{cpu}. 
The algorithm finally returns the chosen mode together with the reasoning (line~\cref{ln:Alg1_8}).

\subsection{Dynamic Function Runtime Adaptation}

The Dynamic Function Runtime Adaptation continuously monitors deployed serverless functions and adjusts their execution mode between \texttt{cpu\_preferred} and \texttt{gpu\_preferred}. To achieve that the Dynamic Function Runtime Adaptation provides adaptive hints to the orchestrator regarding whether a function should execute on CPU or GPU; the orchestrator then handles the scheduling decisions.
It relies on request rate and latency metrics collected from the telemetry, together with configurable thresholds, to decide whether a function should switch its execution environment. This ensures that functions benefit from GPU acceleration when workloads justify it, while avoiding unnecessary GPU usage when the CPU is sufficient.

\begin{figure}
    \centering
    \includegraphics[width=\linewidth]{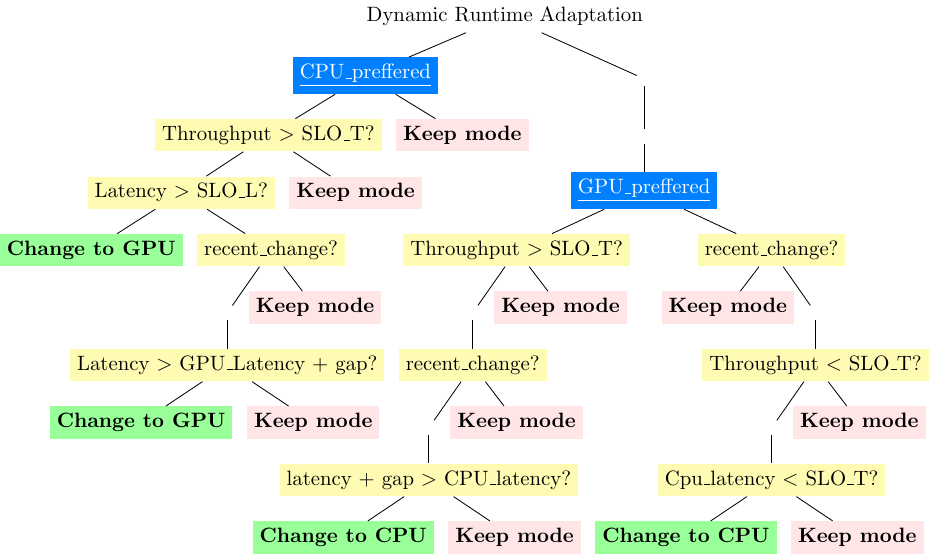}
    \caption{Gaia Dynamic Runtime Function Adaptation}
  \label{fig:gaia-activity}
\end{figure}

\begin{algorithm}[t]
\caption{Dynamic Function Runtime}
\label{alg:dynamic}
\begin{algorithmic}[1]
\Require mode $\in \{\texttt{CPU\_PREF}, \texttt{GPU\_PREF}\}$ 
\Require request rate, latency, thresholds, and saved CPU/GPU latencies
\Ensure Action: change to CPU, change to GPU, or keep mode

\If{mode = \texttt{CPU\_PREF}} \label{ln:dyn1}
  \If{request\_rate $>$ cold\_start\_mitigation\_threshold} \label{ln:dyn2}
    \If{latency $>$ threshold \textbf{or} 
        (recent\_change \textbf{and} latency $>$ saved\_gpu\_latency + gap)} \label{ln:dyn3}
      \State \Return change to GPU \label{ln:dyn4}
    \EndIf
  \EndIf

\ElsIf{mode = \texttt{GPU\_PREF}} \label{ln:dyn5}
  \If{request\_rate $>$ cold\_start\_mitigation\_threshold \textbf{and} recent\_change \textbf{and} 
      (latency + gap $>$ saved\_cpu\_latency)} \label{ln:dyn6}
    \State \Return change to CPU \label{ln:dyn7}
  \EndIf
  \If{request\_rate $<$ threshold \textbf{and} 
      (saved\_cpu\_latency = None \textbf{or} saved\_cpu\_latency $<$ threshold)} \label{ln:dyn8}
    \State \Return change to CPU \label{ln:dyn9}
  \EndIf
\EndIf

\State \Return keep mode (default) \label{ln:dyn10}
\end{algorithmic}
\end{algorithm}

The function runtime adaptation is shown in \cref{fig:gaia-activity} and described \cref{alg:dynamic} as pseudocode. The algorithm starts by checking the current execution mode (line~\ref{ln:dyn1}).  
If the function runs in \texttt{cpu\_preferred} mode, a switch to GPU is only considered when the request rate exceeds the cold start mitigation threshold (line~\ref{ln:dyn2}). 
Within this condition, high latency or a recent change with worse performance compared to the saved GPU latency plus a safety margin (line~\ref{ln:dyn3}) triggers a change to the GPU (line~\ref{ln:dyn4}).  
If the function runs in \texttt{gpu\_preferred} mode (line~\ref{ln:dyn5}), the reevaluator considers switching back to the CPU if two conditions are met.  
First, when the request rate is high, but the observed latency plus a margin is still worse than the saved CPU latency (line~\ref{ln:dyn6}), the mode is switched to CPU (line~\ref{ln:dyn7}).  
Second, when the request rate falls below a lower threshold and CPU performance is acceptable (line~\ref{ln:dyn8}), the function is also switched to CPU (line~\ref{ln:dyn9}).  
If none of these conditions apply, the function keeps its current mode (line~\ref{ln:dyn10}).

\paragraph{Cold Start Mitigation.}  
To prevent mode switches caused by cold start artifacts, Gaia incorporates a request-rate safeguard. Mode changes are only considered if the observed request rate exceeds a configurable threshold, ensuring that outlier latencies from rarely invoked functions do not bias decisions. Additional safeguards, such as a performance-gap margin between CPU and GPU latencies, prevent unnecessary oscillations between modes.  

\section{Prototype Implementation}\label{impl}

The prototype of Gaia is available as open-source\footnote{\url{https://github.com/polaris-slo-cloud/gaia}}. 
Gaia dynamic runtime adaptation is written in Python, while the static code analyzer is written in Go. 

\paragraph{Execution Mode Identifier.}  
The static analyzer inspects Python functions at deployment time. It parses the function source into an abstract syntax tree (AST) and evaluates indicators such as framework imports (e.g., PyTorch, TensorFlow), explicit GPU calls, and tensor operations. 
To integrate Gaia classification into the deployment pipeline, the Knative \texttt{func} CLI was extended with a new \texttt{--deployment-mode} flag (\texttt{auto}, \texttt{cpu}, \texttt{gpu}). When set to \texttt{auto}, the static analyzer is invoked automatically, and its decision is embedded in the Knative Service manifest via annotations and resource limits. This ensures the Kubernetes scheduler correctly places functions on CPU or GPU nodes.  

\paragraph{Dynamic Function Runtime Manager.}  
The dynamic function runtime manager is implemented as a Python service running as a Kubernetes Deployment. It uses the Kubernetes API to inspect Knative Service specifications and queries the telemetry service, i.e, Prometheus, to retrieve runtime metrics. When defined conditions are met, the manager updates the Knative Service manifest to patch resource annotations, allowing Knative to roll out a new revision on the appropriate hardware seamlessly.

\section{Evaluation}\label{evaluation}

To evaluate Gaia, we designed a set of workloads representing common and contrasting workload types. We aim to evaluate our framework across different execution modes (CPU vs.~GPU) and usage scenarios, highlighting strengths and limitations. Each workload was deployed and executed sequentially to ensure isolated measurements. Our goal is to identify the best execution mode for workloads with different characteristics by balancing performance and cost tradeoffs.

\paragraph{Experimental Workloads} We evaluate the framework with four representative workloads that stress different aspects of CPU/GPU execution. Our workloads are:
\begin{enumerate*}
    \item Matrix multiplication models, an intensive computing task where complexity grows with input size, exposing the decision runtime’s ability to adapt execution modes.
    \item Image classification with ResNet-18~\cite{he_deep_2015}, and 
    \item Large language model (LLM) inference with TinyLlama~\cite{tinyLlama} represents realistic AI tasks that benefit from GPU acceleration, ranging from moderate to highly demanding inference.
    \item Idle wait function, parameterized by \texttt{wait\_time}, provides a degenerate case with no useful computation, serving as a lower bound for performance comparison.
\end{enumerate*}
Although our illustrative scenario (\cref{subsec:scenario}) highlights deforestation within the 3D Compute Continuum, Gaia focuses on hardware acceleration across varying workloads, independent of specific application domains. Therefore, to evaluate Gaia, we utilize controlled and representative AI workloads, including compute-intensive tasks, small AI models, and Large Language Models (LLMs), which reflect the computational and memory characteristics of typical real-world use cases, including the scenarios we illustrate. Thus, our experimental design isolates Gaia's contributions and facilitates a reproducible, hardware-agnostic evaluation of its decision-making logic.

\paragraph{Metrics and Baselines} For all experimental workloads, we collect latency and Costs. The results are collected for CPU, GPU, and Gaia, and the prices are based on Microsoft Azure pricing model~\cite{azurePrizing}. Our goal is to evaluate for each use case the performance and cost trade-offs of executing on CPU and GPU.

\subsection{Experimental Setup}

To evaluate Gaia, we deployed a MicroK8s~\cite{microk8s} cluster with three nodes with different hardware configurations, to mimic the heterogeneous environments of the 3D Compute Continuum. One node with 4vCPUs, no GPU, and 8GB RAM, and two worker nodes. The first worker node has 8vCPUs, 32 GB RAM, and no GPU, and a worker with 16GB, 64 GB RAM, and a GPU NVIDIA GeForce RTX 3090 24GB.
 Each node runs Ubuntu Server for ARM 24.04 LTS. The experimental workflow is deployed as Knative~\cite{knative} services. To avoid bias in the results, all experiments were executed 5 times, and the results shown are the corresponding mean values.

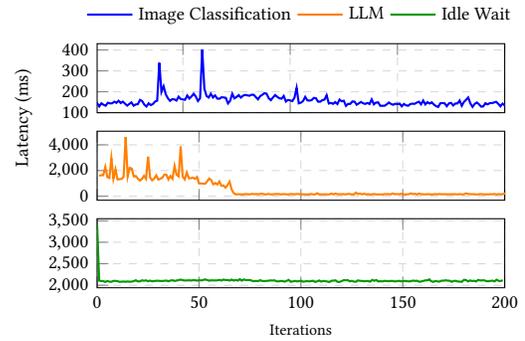
\begin{figure}[t]  
\centering
    \begin{tikzpicture}
            \begin{groupplot}[
              group style={group size=1 by 3, vertical sep=7pt},
              width=7cm, height=2.5cm,
              axis lines=box, ytick pos=left,
              yticklabel style={font=\footnotesize},
              xticklabel style={font=\footnotesize},  
              grid=major, grid style={dashed,gray!30},
              ylabel style={yshift=5pt,font=\footnotesize},
              ylabel={Latency (ms)},
            ]
            
            % ----- Row 1 (visible y-axis + label) -----
            \nextgroupplot[
              ylabel style={font=\footnotesize, xshift=-15pt},
              legend style={at={(0.5,1.7)}, anchor=north, draw=none, font=\footnotesize, legend columns=3},
              xmin=10,
              xmax=200,
              xticklabels={}
            ]
            \addplot[draw=blue, thick, mark=none] table[x=Index,y=response_time,col sep=comma,mark=none] {experiments/2_image/data/response_time.csv};

            \addlegendentry{Image Classification}
            \addlegendentry{LLM}
            \addlegendimage{line legend, draw=orange, thick}
            \addlegendentry{Idle Wait}
            \addlegendimage{line legend, draw=green!60!black, thick}
            % ----- Row 2 (visible y-axis ticks, no ylabel text) -----
            \nextgroupplot[axis lines=box, ytick pos=left, 
            xmin=0,
            xmax=200,
            ylabel={},
            xticklabels={},
            xlabel style={font=\scriptsize}]
            \addplot[draw=orange, thick, mark=none]  table[x=Index,y=response_time,col sep=comma,mark=none,draw=orange] {experiments/3_llm/data/response_time.csv};

            \nextgroupplot[axis lines=box, ytick pos=left, 
            ylabel={},
            xlabel={Iterations},
            xmin=0,
            xmax=200,
            xlabel style={font=\scriptsize}]
            
            \addplot[draw=green!60!black, thick, mark=none]  table[x=Index,y=response_time,col sep=comma,draw=green!60!black,mark=none] {experiments/4_idle/data/response_time.csv};
            
            \end{groupplot}
            
        \end{tikzpicture}
    \caption{Serverless Function Execution Latency with Dynamic Function Runtime Reconfiguration}
    \label{fig:latency}
\end{figure}

\begin{figure*}[t]
    % Propagate - T_Total Performance Latency and Throughput 
    % Read Distance
    \begin{subfigure}{0.19\linewidth}
      \begin{tikzpicture}
        \begin{axis}[
          xlabel={Matrix Size}, % change if you want units
          ylabel={Milliseconds},
          ylabel style={yshift=-2pt,font=\footnotesize},
          xticklabel style={font=\footnotesize},
          xlabel style={font=\footnotesize},
          yticklabel style={font=\footnotesize},
          width=4cm, height=3.5cm,
          grid=major, grid style={dashed,gray!30},
          mark options={solid},
          legend style={at={(3,1.1)},anchor=south,draw=none,legend columns=-1},
        ]
          % the others:
          \addplot+[no marks, color=orange, thick] table[x=Index, y={CPU_response_time},    col sep=comma] {experiments/1_matrix/data/response_time.csv};
          \addplot+[no marks, color=green!60!black, thick] table[x=Index, y={GPU_response_time}, col sep=comma] {experiments/1_matrix/data/response_time.csv};
          \addplot+[no marks, color=blue, thick]
          table[x=Index, y=AUTO_response_time, col sep=comma]
          {experiments/1_matrix/data/response_time.csv};
          \legend{CPU, GPU, Gaia}
        \end{axis}
      \end{tikzpicture}
      \caption{Latency}
      \label{fig:matrix-multi-boxplot}
    \end{subfigure}
       \begin{subfigure}{0.19\linewidth}
        \begin{tikzpicture}
            \begin{axis}[               
                xlabel={Matrix Size},
                ylabel style={yshift=-2pt,font=\footnotesize},
                ylabel={Costs (USD)},
                xticklabel style={font=\footnotesize},  
                xlabel style={font=\footnotesize},
                yticklabel style={font=\footnotesize}, 
                width=4cm,                     
                height=3.5cm,
                grid=major,
                grid style={dashed,gray!30},
                mark options={solid}
            ]
           
            \addplot+[no marks, color=green!60!black, thick] table[x=Index, y={GPU_cost}, col sep=comma] {experiments/1_matrix/data/response_time.csv};
          \addplot+[no marks, color=orange, thick] table[x=Index, y={CPU_cost},    col sep=comma] {experiments/1_matrix/data/response_time.csv};
          \addplot+[no marks, color=blue, thick]
          table[x=Index, y=AUTO_cost, col sep=comma]
          {experiments/1_matrix/data/response_time.csv};
            
            \end{axis}
        \end{tikzpicture}
        \caption{Costs}
        \label{fig:matrix-multi-cost}
    \end{subfigure}
    % MIG
    \begin{subfigure}{0.19\linewidth}
        \begin{tikzpicture}
            \begin{axis}[               
                xlabel={Time (sec)},
                ylabel style={yshift=-2pt,font=\footnotesize},
                xticklabel style={font=\footnotesize},  
                xlabel style={font=\footnotesize},
                yticklabel style={font=\footnotesize}, 
                ylabel={CPU Utilization (Cores)},
                width=4cm,                    
                height=3.5cm,
                grid=major,
                xmax=300,
                grid style={dashed,gray!30},
                mark options={solid}
            ]
          \addplot+[no marks, color=green!60!black, thick] table[x=Index, y={GPU}, col sep=comma] {experiments/1_matrix/data/cpu.csv};
          \addplot+[no marks, color=orange, thick] table[x=Index, y={CPU},    col sep=comma] {experiments/1_matrix/data/cpu.csv};
          \addplot+[no marks, color=blue, thick]
          table[x=Index, y=AUTO-cpu-revision, col sep=comma]
          {experiments/1_matrix/data/cpu.csv};
            \end{axis}
        \end{tikzpicture}
        \caption{CPU}
        \label{fig:matrix-multi-cpu}
    \end{subfigure}
    % Throughput
    \begin{subfigure}{0.19\linewidth}
        \begin{tikzpicture}
            \begin{axis}[               
                xlabel={Time (sec)},
                ylabel style={yshift=-2pt,font=\footnotesize},
                ylabel={},
                xticklabel style={font=\footnotesize},  
                xlabel style={font=\footnotesize},
                yticklabel style={font=\footnotesize}, 
                ylabel={Ram Utilization (GB)},
                width=4cm,                      
                height=3.5cm,
                grid=major,
                xmax=300,
                grid style={dashed,gray!30},
                mark options={solid}
            ]
            \addplot+[no marks, color=green!60!black, thick] table[x=Index, y={GPU}, col sep=comma] {experiments/1_matrix/data/ram.csv};
          \addplot+[no marks, color=orange, thick] table[x=Index, y={CPU},    col sep=comma] {experiments/1_matrix/data/ram.csv};
          \addplot+[no marks, color=blue, thick]
          table[x=Index, y=AUTO-cpu-revision, col sep=comma]
          {experiments/1_matrix/data/ram.csv};
            \end{axis}
        \end{tikzpicture}
        \caption{RAM}
        \label{fig:matrix-multi-ram}
    \end{subfigure}
    % Throughput
    \begin{subfigure}{0.19\linewidth}
        \begin{tikzpicture}
            \begin{axis}[               
                xlabel={Time (sec)},
                ylabel style={yshift=-2pt,font=\footnotesize},
                ylabel={},
                xticklabel style={font=\footnotesize},  
                xlabel style={font=\footnotesize},
                yticklabel style={font=\footnotesize}, 
                ylabel={GPU Utilization (\%)},
                width=4cm,                      
                height=3.5cm,
                xmax=300,
                grid=major,
                grid style={dashed,gray!30},
                mark options={solid}
            ]
            \addplot+[no marks, color=green!60!black, thick] table[x=Index, y={GPU}, col sep=comma] {experiments/1_matrix/data/gpu.csv};
          \addplot+[no marks, color=blue, thick]
          table[x=Index, y=AUTO, col sep=comma]
          {experiments/1_matrix/data/gpu.csv};
            \end{axis}
        \end{tikzpicture}
        \caption{GPU}
        \label{fig:matrix-multi-gpu}
    \end{subfigure}
    \caption{Matrix multiplication: Gaia promotes to GPU once the SLO is hit, producing a clear step-down in latency and cost relative to CPU.}
    \label{fig:maxtrix_performance}
\end{figure*}
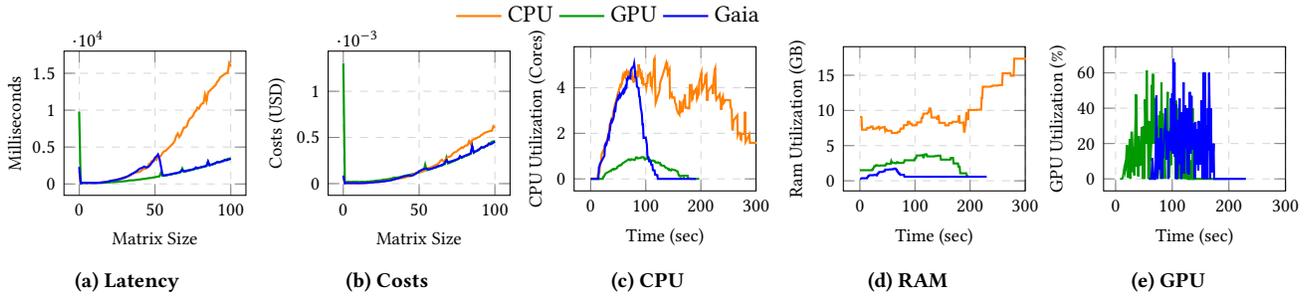

\begin{figure*}[t]
    % Propagate - T_Total Performance Latency and Throughput 
    % Read Distance
    \begin{subfigure}{0.17\linewidth}
      \begin{tikzpicture}
          \begin{axis}[
            boxplot/draw direction=y,
            box plot width=3mm,
            width=3.5cm, height=3.5cm,
            ymajorgrids=true, grid style=dashed,
            xtick={1,2,3},
            xlabel={~},
            xticklabels={CPU,GPU,Gaia},
            xticklabel style={font=\scriptsize},
            ylabel style={yshift=-2pt,font=\footnotesize},
            enlarge x limits=0.1,
            ymode=log,
            ylabel={Latency (ms)}
          ]
        
              \addplot+[
                boxplot prepared={
                  median=1417.460680,
                  upper quartile=1667.139030,
                  lower quartile=1326.127434,
                  upper whisker=2147.633886,
                  lower whisker=1226.755190,
                  draw position=1
                },
                fill=orange!40,       % box fill color
                draw=orange!80,       % outline color
              ] coordinates {(1,0)};
            
              \addplot+[
                boxplot prepared={
                  median=157.382250,
                  upper quartile=172.411156,
                  lower quartile=146.947408,
                  upper whisker=210.494232,
                  lower whisker=135.144567,
                  draw position=2
                },
                fill=green!60!black!40,       % box fill color
                draw=green!60!black!80,       % outline color
              ] coordinates {(2,0)};
            
              \addplot+[
                boxplot prepared={
                  median=155.229330,
                  upper quartile=178.904939,
                  lower quartile=144.543910,
                  upper whisker=222.838306,
                  lower whisker=135.617971,
                  draw position=3
                },
                fill=blue!40,       % box fill color
                draw=blue!80,       % outline color
              ] coordinates {(3,0)};
          \end{axis}
        \end{tikzpicture}
        
      \caption{Latency}
      \label{fig:llm-boxplot}
    \end{subfigure}
       \begin{subfigure}{0.17\linewidth}
        \begin{tikzpicture}
            \begin{axis}[
                ybar,
                symbolic x coords={CPU, GPU, Gaia},
                xtick=data,
                ymin=0,
                ylabel={Costs (USD)},
                xlabel={~},
                xticklabel style={font=\footnotesize},  
                ylabel style={font=\footnotesize},
                yticklabel style={font=\footnotesize}, 
                bar width=10pt,
                width=3.5cm,         
                height=3.5cm,
                enlarge x limits=0.2,
                tick align=inside
            ]
                \addplot[fill=blue!50, error bars/.cd, y dir=both,
                    y explicit, error mark=none ] coordinates {
                    (CPU,0.032)
                    (GPU,0.019)
                    (Gaia,0.019)
                };

            \end{axis}
        \end{tikzpicture}
        \caption{Costs}
        \label{fig:llm-cost}
    \end{subfigure}
    % MIG
    \begin{subfigure}{0.21\linewidth}
        \begin{tikzpicture}
            \begin{axis}[               
                ylabel style={yshift=-2pt,font=\footnotesize},
                xticklabel style={font=\footnotesize},  
                xlabel style={font=\footnotesize},
                yticklabel style={font=\footnotesize}, 
                ylabel={CPU Utilization (Cores)},
                xlabel={Time (sec)},
                width=4cm,                    
                height=3.3cm,
                grid=major,
                xmax=300,
                grid style={dashed,gray!30},
                mark options={solid},
                legend style={at={(0.5,1.2)},anchor=south,draw=none,legend columns=-1},
            ]
          \addplot+[no marks, color=orange, thick] table[x=Index, y={CPU},    col sep=comma] {experiments/3_llm/data/cpu.csv};
          \addplot+[no marks, color=green!60!black, thick] table[x=Index, y={GPU}, col sep=comma] {experiments/3_llm/data/cpu.csv};
          \addplot+[no marks, color=blue, thick]
          table[x=Index, y=AUTO-cpu-revision, col sep=comma]
          {experiments/3_llm/data/cpu.csv};

          \legend{CPU, GPU, Gaia}
            \end{axis}
        \end{tikzpicture}
        \caption{CPU}
        \label{fig:llm-cpu}
    \end{subfigure}
    % Throughput
    \begin{subfigure}{0.21\linewidth}
        \begin{tikzpicture}
            \begin{axis}[               
                xlabel={Time (sec)},
                ylabel style={yshift=-5pt,font=\footnotesize},
                ylabel={},
                xticklabel style={font=\footnotesize},  
                xlabel style={font=\footnotesize},
                yticklabel style={font=\footnotesize}, 
                ylabel={Ram Utilization (GB)},
                width=4cm,                      
                height=3.3cm,
                grid=major,
                xmax=300,
                grid style={dashed,gray!30},
                mark options={solid}
            ]
            
          \addplot+[no marks, color=orange, thick] table[x=Index, y={CPU},    col sep=comma] {experiments/3_llm/data/ram.csv};
          \addplot+[no marks, color=green!60!black, thick] table[x=Index, y={GPU}, col sep=comma] {experiments/3_llm/data/ram.csv};
          \addplot+[no marks, color=blue, thick]
          table[x=Index, y=AUTO-cpu-revision, col sep=comma]
          {experiments/3_llm/data/ram.csv};
           %\legend{CPU, GPU, Gaia}
            \end{axis}
        \end{tikzpicture}
        \caption{RAM}
        \label{fig:llm-ram}
    \end{subfigure}
    % Throughput
    \begin{subfigure}{0.21\linewidth}
        \begin{tikzpicture}
            \begin{axis}[               
                xlabel={Time (sec)},
                ylabel style={yshift=-5pt,font=\footnotesize},
                ylabel={},
                xticklabel style={font=\footnotesize},  
                xlabel style={font=\footnotesize},
                yticklabel style={font=\footnotesize}, 
                ylabel={GPU Utilization (\%)},
                width=4cm,                      
                height=3.3cm,
                xmax=300,
                grid=major,
                grid style={dashed,gray!30},
                mark options={solid}
            ]
            \addplot+[no marks, color=green!60!black, thick] table[x=Index, y={GPU}, col sep=comma] {experiments/3_llm/data/gpu.csv};
          \addplot+[no marks, color=blue, thick]
          table[x=Index, y=AUTO, col sep=comma]
          {experiments/3_llm/data/gpu.csv};
            \end{axis}
        \end{tikzpicture}
        \caption{GPU}
        \label{fig:llm-gpu}
    \end{subfigure}
    \caption{LLM: The initial CPU phase triggers promotion; latency stabilizes at the accelerated tier at a lower cost after switching to GPU}
    \label{fig:performance}
\end{figure*}
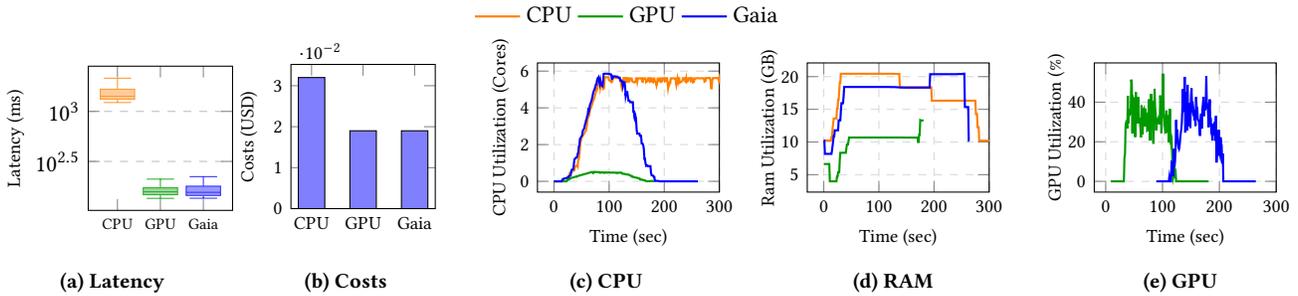

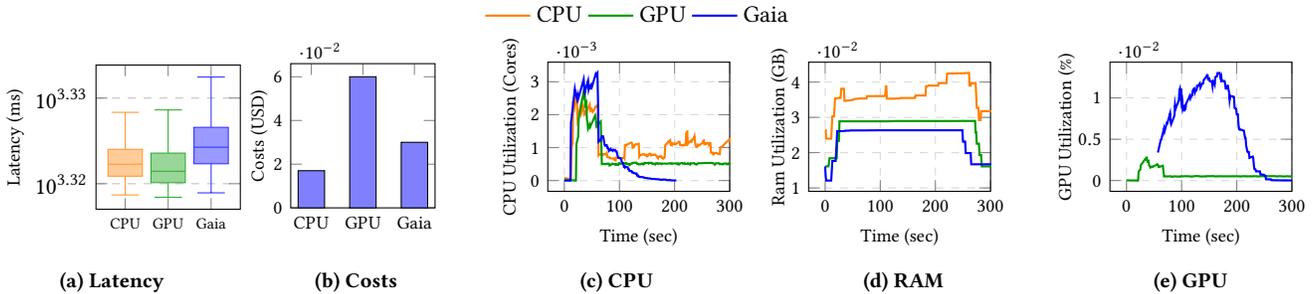
\begin{figure*}[t]
    % Propagate - T_Total Performance Latency and Throughput 
    % Read Distance
    \begin{subfigure}{0.17\linewidth}
      \begin{tikzpicture}
          \begin{axis}[
            boxplot/draw direction=y,
            box plot width=3mm,
            width=3.5cm, height=3.5cm,
            ymajorgrids=true, grid style=dashed,
            xtick={1,2,3},
            xlabel={~},
            xticklabels={CPU,GPU,Gaia},
            xticklabel style={font=\scriptsize},
            ylabel style={yshift=-2pt,font=\footnotesize},
            enlarge x limits=0.1,
            ymode=log,
            ylabel={Latency (ms)}
          ]
        
              \addplot+[
                boxplot prepared={
                  median=2091.532850, upper quartile=2099.936461, lower quartile=2084.772205, upper whisker=2120.855331, lower whisker=2074.326992,
                  draw position=1
                },
                fill=orange!40,       % box fill color
                draw=orange!80,       % outline color
              ] coordinates {(1,0)};
            
              \addplot+[
                boxplot prepared={
                  median=2087.567520, upper quartile=2097.758055, lower quartile=2081.152987, upper whisker=2122.289753, lower whisker=2072.992992,
                  draw position=2
                },
                fill=green!60!black!40,       % box fill color
                draw=green!60!black!80,       % outline color
              ] coordinates {(2,0)};
            
              \addplot+[
                boxplot prepared={
                  median=2101.088190, upper quartile=2112.379575, lower quartile=2091.780210, upper whisker=2141.105652, lower whisker=2075.471449,
                  draw position=3
                },
                fill=blue!40,       % box fill color
                draw=blue!80,       % outline color
              ] coordinates {(3,0)};
        
          \end{axis}
        \end{tikzpicture}
        
      \caption{Latency}
      \label{fig:idle-latency}
    \end{subfigure}
    \hspace{0.2em}
       \begin{subfigure}{0.17\linewidth}
        \begin{tikzpicture}
            \begin{axis}[
                ybar,
                symbolic x coords={CPU, GPU, Gaia},
                xtick=data,
                ymin=0,
                ylabel={Costs (USD)},
                xlabel={~},
                ylabel style={yshift=-4pt,font=\footnotesize},
                ylabel style={font=\footnotesize},
                yticklabel style={font=\footnotesize}, 
                xticklabel style={font=\footnotesize},  
                bar width=10pt,
                width=3.5cm,         
                height=3.5cm,
                enlarge x limits=0.2,
                tick align=inside
            ]
                \addplot[fill=blue!50, error bars/.cd, y dir=both,
                    y explicit, error mark=none ] coordinates {
                    (CPU,0.017)
                    (GPU,0.060)
                    (Gaia,0.03)
                };

            \end{axis}
        \end{tikzpicture}
        \caption{Costs}
        \label{fig:idle-cost}
    \end{subfigure}
    \begin{subfigure}{0.21\linewidth}
        \begin{tikzpicture}
            \begin{axis}[               
                ylabel style={yshift=-2pt,font=\footnotesize},
                xticklabel style={font=\footnotesize},  
                xlabel style={font=\footnotesize},
                yticklabel style={font=\footnotesize}, 
                ylabel={CPU Utilization (Cores)},
                xlabel={Time (sec)},
                width=4cm,                    
                height=3.3cm,
                grid=major,
                xmax=300,
                grid style={dashed,gray!30},
                mark options={solid},
                legend style={at={(0.5,1.2)},anchor=south,draw=none,legend columns=-1},
            ]
          
          \addplot+[no marks, color=orange, thick] table[x=Index, y={CPU},    col sep=comma] {experiments/4_idle/data/cpu.csv};
          \addplot+[no marks, color=green!60!black, thick] table[x=Index, y={GPU}, col sep=comma] {experiments/4_idle/data/cpu.csv};
          \addplot+[no marks, color=blue, thick]
          table[x=Index, y=AUTO-first-cpu-revision, col sep=comma]
          {experiments/4_idle/data/cpu.csv};
          \legend{CPU, GPU, Gaia}
            \end{axis}
        \end{tikzpicture}
        \caption{CPU}
        \label{fig:idle-cpu}
    \end{subfigure}
    % Throughput
    \begin{subfigure}{0.21\linewidth}
        \begin{tikzpicture}
            \begin{axis}[               
                xlabel={Time (sec)},
                ylabel style={yshift=-5pt,font=\footnotesize},
                ylabel={},
                xticklabel style={font=\footnotesize},  
                xlabel style={font=\footnotesize},
                yticklabel style={font=\footnotesize}, 
                ylabel={Ram Utilization (GB)},
                width=4cm,                      
                height=3.3cm,
                grid=major,
                xmax=300,
                grid style={dashed,gray!30},
                mark options={solid}
            ]
            \addplot+[no marks, color=green!60!black, thick] table[x=Index, y={GPU}, col sep=comma] {experiments/4_idle/data/ram.csv};
          \addplot+[no marks, color=orange, thick] table[x=Index, y={CPU},    col sep=comma] {experiments/4_idle/data/ram.csv};
          \addplot+[no marks, color=blue, thick]
          table[x=Index, y=AUTO-first-cpu-revision, col sep=comma]
          {experiments/4_idle/data/ram.csv};
            \end{axis}
        \end{tikzpicture}
        \caption{RAM}
        \label{fig:idle-ram}
    \end{subfigure}
    % Throughput
    \begin{subfigure}{0.21\linewidth}
        \begin{tikzpicture}
            \begin{axis}[               
                xlabel={Time (sec)},
                ylabel style={yshift=-5pt,font=\footnotesize},
                ylabel={},
                xticklabel style={font=\footnotesize},  
                xlabel style={font=\footnotesize},
                yticklabel style={font=\footnotesize}, 
                ylabel={GPU Utilization (\%)},
                width=4cm,                      
                height=3.3cm,
                xmax=300,
                grid=major,
                grid style={dashed,gray!30},
                mark options={solid}
            ]
            \addplot+[no marks, color=green!60!black, thick] table[x=Index, y={GPU}, col sep=comma] {experiments/4_idle/data/cpu.csv};
          \addplot+[no marks, color=blue, thick]
          table[x=Index, y=AUTO-gpu-revision, col sep=comma]
          {experiments/4_idle/data/cpu.csv};
            \end{axis}
        \end{tikzpicture}
        \caption{GPU}
        \label{fig:idle-gpu}
    \end{subfigure}
    \caption{Idle Wait: after GPU promotion and no latency reduction, Gaia downgrades to CPU again.}
    \label{fig:idle-response}
\end{figure*}

\subsection{Experimental Results}

\subsubsection{Overall latency}

\emph{\cref{fig:latency} shows that Gaia reduces latency up to 95\% once it promotes the backend from CPU to GPU.}

\emph{Image Classification:} Gaia sustains a median response time of 145 ms with rare spikes up to 403ms. However, the latency spikes are not sufficient, and Image Classification runs entirely on CPU.

\emph{LLM:} In this experiment, Gaia shows the characteristic two-regime curve: an initial steady response times predominantly in 1.3–2.3,s (occasional outliers up to 4.6,s), followed by a sharp step-down to a stable band around 140–200 ms. Once SLOs are violated, Gaia’s runtime policy switches from a CPU-bound path to an accelerated path (GPU or equivalent fast tier), thereby significantly reducing latency. 

\emph{Idle Wait:} Gaia first changes the execution mode to GPU because of the high latency. After that, the algorithm detects no significant difference latency and thus changes the execution mode back to CPU, maintaining a latency of near 2 seconds 2s.

\subsubsection{Matrix Multiplication Results} 

\cref{fig:maxtrix_performance} shows that with increasing matrix size, Gaia execution mode identifier returns CPU, which explains the initial high latency for GaiaC before the promotion to GPU rises.

\emph{Latency:} shows that all modes behave similarly for small matrices. After the SLO is reached, Gaia switches to a GPU-back, achieving a clear step-down in response time compared to the CPU. The boxplot in \cref{fig:matrix-multi-boxplot} captures this distributional shift: CPU develops long tails as size increases, while Gaia collapses latency after promotion.

\emph{Cost:} As execution time grows with matrix size, CPU becomes the most expensive option (\cref{fig:matrix-multi-cost}). Gaia tracks CPU-like costs early on, then, after promotion, follows the lower GPU cost curve because requests are completed faster. GPU bears the largest cost in the beginning due to cold-start cost, but is competitive once warmed.

\emph{CPU/RAM usage:} \cref{fig:matrix-multi-cpu} and \cref{fig:matrix-multi-ram} show that GPU-backed execution uses significantly less CPU and RAM than the CPU function for the same workload window. The two orange vertical markers denote the Gaia switch points. \cref{fig:matrix-multi-gpu} shows that Gaia engages the GPU later than GPU-mode: the latter runs on the GPU immediately (after its large cold start), while Gaia promotes only when SLOs are violated, minimizing time spent on expensive or unnecessary acceleration.

\subsubsection{LLM Inference Results}

For the prompt (``What is the capital of France?''), Gaia identified CPU initially; as soon as the running latency violated the SLO, Gaia promoted to GPU. Post-promotion, Gaia behaves indistinguishably from GPU mode in both latency and cost, while avoiding the large GPU cold-start penalty.

\emph{Latency:} The boxplot in \cref{fig:llm-boxplot} shows that once promoted, Gaia aligns with the GPU, while the remaining outliers are attributable to the pre-promotion CPU phase, and CPU exhibits higher latency overall.

\emph{Cost:} \cref{fig:llm-cost} shows that using the measured totals (CPU: 0.03206, GPU: 0.01914, Gaia: 0.01910), Gaia and GPU are 40\% cheaper than CPU. Gaia and GPU are identical following the direct latency collapse after promotion (\cref{fig:llm-boxplot}).

\emph{Resource usage:} On CPU, utilization is high; on Gaia after promotion to GPU, CPU load drops and GPU utilization rises to serve inference efficiently (\cref{fig:llm-cpu,fig:llm-gpu}). RAM (\cref{fig:llm-ram}) per request stabilizes/declines due to shorter service time on the GPU.

\subsubsection{Idle Function Results}
\cref{fig:idle-response} shows that Gaia briefly promotes to the GPU on high latency, finds no improvement, and returns to the CPU. Net effect: CPU-like latency and cost with a short GPU detour.

\emph{Latency:} High initial latency triggers a promotion to GPU, but because the task is not GPU-amenable, latency does not drop; thus, Gaia demotes back to CPU (\cref{fig:idle-latency}). The boxplot confirms Gaia aligns with CPU, aside from the short GPU window.

\emph{Cost:} The cost curve shows a small hump exactly during the GPU interval, then converges to the CPU cost trajectory (\cref{fig:idle-cost}). 

\emph{Resource Usage:} Resource traces (\cref{fig:idle-cpu,fig:idle-ram}) remain similar across modes; no sustained CPU or RAM relief appears on the GPU because the workload spends time waiting rather than computing.

\section{Related Work}\label{related_work}
In this section, we present an analysis of the state-of-the-art challenges and how they differentiate from Gaia in the 3D Compute Continuum and GPU-enabled serverless computing. 

\subsection{Serverless in the 3D Compute Continuum}

HyperDrive \cite{HyperDrive2024} proposes an architecture for serverless computing in the 3D Continuum, featuring a function scheduler that accounts for the dynamic and heterogeneous characteristics of the environment. While HyperDrive incorporates environment characteristics such as node vicinity, thermal conditions, and power constraints, it does not adapt during runtime to workload-specific characteristics. 
OrbitFaaS \cite{OrbiFaaS2025} introduces a serverless platform for EO data processing by leveraging in-situ computing capacity. OrbitFaaS introduces orbital models and communication abstractions that enable optimized data processing and AI workloads in the 3D Continuum. Komet \cite{Komet2024} proposes a LEO-tailored serverless platform that seamlessly replicates function data in the satellite orbit trajectory. Databelt \cite{Databelt} enables stateful serverless function execution by proactively placing function state along the path of the workflow execution.
However, OrbitFaas, Komet, and Databelt focus on the data management and exchange between functions and do not provide workload-aware execution, which impacts the performance of the function. 
ML-SDN \cite{ResilientFaas} proposes a serverless software-defined networking (SDN) architecture that dynamically orchestrates communications and computation resources to meet diverse 6G service-level agreements (SLAs). By dynamically adjusting workloads, ML-SDN improves end-to-end learning performance. Krios \cite{Krios2024} introduces scheduling abstractions for serverless functions that enable function placement in a specific geographic region. Krios software-defined LEO zones leverage satellite path prediction models to proactively predict serverless functions handovers. 
However, ML-SDN and Krios primarily focus on network and control-plane orchestration and do not provide function abstraction that allows the platform to seamlessly decide on hardware acceleration. Cosmos \cite{cosmos} proposes a performance-cost tradeoff model for function placement in the Edge-Cloud-Space 3D Continuum. However, Cosmos does not address specific node requirements.

Although the discussed state-of-the-art enables serverless execution in the 3D Compute Continuum, they either focus on network or specific orchestration components that do not consider heterogeneous workload characteristics of serverless functions. In contrast, Gaia seamlessly identifies hardware acceleration and introduces runtime workload adaptation to meet node and user-defined performance requirements.

\subsection{GPU-enabled Serverless Computing}

Several approaches enable GPU support in serverless platforms. Kim et al.~\cite{kim_gpu_2018}, the OSCAR Extension~\cite{naranjo_accelerated_2020}, and KaaS~\cite{pemberton_kernel-as--service_2022} require developers to explicitly specify whether a function executes on a CPU or GPU. Several commercial platforms, including RunPod~\cite{runpod}, Cerebrium~\cite{cerebrium}, and Beam Cloud~\cite{beam-cloud}, offer GPU sharing but do not include automated decision procedures for CPU versus GPU execution.  
While this method provides full control over resource allocation and ensures efficient GPU use when configured properly, it increases development complexity and risks inefficient resource utilization. 
Dynamic approaches such as Llama~\cite{romero_llama_2021} and DGSF~\cite{fingler_dgsf_2022} automatically determine the execution device before runtime based on resource availability or user-defined SLOs such as performance, cost. While they allow efficient resource usage and adaptability to workload changes, they also introduce additional decision-making overhead and increase system complexity. 
Protean \cite{bhasi_towards_2024} proposes a GPU-enabled serverless framework that leverages new-generation NVIDIA GPU capabilities, such as Multi-Instance GPU (MIG) and Multi-Process Service (MPS), to improve SLO compliance and reduce costs. Although Protean avoids CPU-related overhead and maximizes GPU performance, it runs all tasks exclusively on GPUs, sacrificing flexibility and may incur unnecessary costs when workloads do not require GPU acceleration. 

Although existing work advances GPU-enabled serverless computing, it relies on manual developer intervention or performs static device selection prior to execution. In contrast, Gaia combines pre-execution decisions with runtime workload adaptation, enabling the platform to seamlessly adjust hardware acceleration during execution according to performance and cost requirements.

\section{Conclusion}\label{conclusion}
In this paper, we presented \emph{Gaia}, a GPU-as-a-Service model and architecture that shifts device selection from developers to the platform. Gaia combines an Execution Mode Identifier, which assigns an initial hardware acceleration mode from source code, and a Dynamic Function Runtime that revisits this choice based on SLOs. Our experimental results show that Gaia promotes accelerable workloads and demotes non-accelerable ones. Across matrix multiplication, Image classification, LLM inference, and an idle baseline, reducing end-to-end latency by up to 95\%

In future work, we will enhance dynamic runtime decisions by incorporating predictive, learning-based policies and workflow-level objectives.  Additionally, we will integrate the dynamics of the 3D Compute Continuum, including LEO handovers, power and thermal constraints, and intermittent links, into our placement and mode decisions.

\begin{acks}
This work has received funding from the Austrian Internet Stiftung under the NetIdee project LEO Trek (ID~7442).
Partly funded by EU Horizon Europe under GA no. 101070186 (TEADAL) and GA no. 101192912 (NexaSphere). 
\end{acks}

\bibliographystyle{ACM-Reference-Format}
\bibliography{references}

\balance
\appendix

\end{document}